\documentclass{jpp}
\usepackage{graphicx}
\usepackage{epstopdf, epsfig}
\usepackage{hyperref}

\newcommand{\Ar}{\mathcal{A}_{\rm R}}
\newcommand{\Hr}{H_{\rm R}}
\newcommand{\Ab}{\boldsymbol{A}}
\newcommand{\Ap}{\boldsymbol{A}_{\rm p}}
\newcommand{\Bb}{\boldsymbol{B}}

\newcommand{\Bp}{\boldsymbol{B}_{\rm p}}

\newcommand{\nb}{\hat{\boldsymbol{n}}}
\newcommand{\lb}{\boldsymbol{l}}

\newcommand{\xb}{\boldsymbol{x}}
\newcommand{\yb}{\boldsymbol{y}}
\newcommand{\zb}{\boldsymbol{z}}

\usepackage{xcolor}

\shorttitle{Relative field-line helicity}
\shortauthor{A. R. Yeates and M. H. Page}

\title{Relative field-line helicity in bounded domains}

\author{Anthony R. Yeates
  \corresp{\email{anthony.yeates@durham.ac.uk}},
  Marcus H. Page}

\affiliation{Department of Mathematical Sciences, Durham University, South Road, Durham DH1 3LE, UK}

\begin{document}

\maketitle

\begin{abstract}
Models for astrophysical plasmas often have magnetic field lines that leave the boundary rather than closing within the computational domain. Thus, the relative magnetic helicity is frequently used in place of the usual magnetic helicity, so as to restore gauge invariance. We show how to decompose the relative helicity into a relative field-line helicity that is an ideal-magnetohydrodynamic invariant for each individual magnetic field line, and vanishes along any field line where the original field matches the reference field. Physically, this relative field-line helicity is a magnetic flux, whose specific definition depends on the gauge of the reference vector potential on the boundary. We propose a particular ``minimal'' gauge that depends only on the reference field and minimises this boundary contribution, so as to reveal topological information about the original magnetic field. We illustrate the effect of different gauge choices using the Low-Lou and Titov-D\'emoulin models of solar active regions. Our numerical code to compute appropriate vector potentials and relative field-line helicity in Cartesian domains is open source and freely available.
\end{abstract}

\section{Introduction} \label{sec:intro}

Magnetic helicity, $h(V)$, has been known to be an integral invariant of ideal magnetohydrodynamics (MHD) since \citet{1958PNAS...44..489W}, and its elegant physical interpretation as the average pairwise linking of magnetic flux tubes (or magnetic field lines)   since  \cite{1969JFM....35..117M}. Its practical importance is largely due to its robustness even in the presence of finite resistivity \citep[see, e.g.,][]{1984GApFD..30...79B,1988JPlPh..40..263B}.
However, the linking interpretation requires the field lines to be either closed curves, or to be ergodic within the domain \citep{arnold1986}. 
In astrophysical situations, the magnetic field is typically not confined to a finite volume, so that field lines will typically leave through the boundary of any finite computational domain.

The seminal work of \citet{1984JFM...147..133B} showed how to define a gauge-invariant version of $h(V)$ even when the domain is not magnetically closed, by subtracting the helicity of an appropriate reference magnetic field. This so-called relative magnetic helicity, $\Hr$, effectively measures the average linking with respect to that of the reference field. Typically one chooses a reference field with as little self-linking as possible, so that the desired information about the structure of the original field itself is revealed. Usually, this is done by fixing the reference field to the unique current-free/potential magnetic field that minimises magnetic energy for the given boundary conditions. Like $h(V)$, the relative helicity $\Hr$ is invariant under an ideal-MHD evolution that vanishes on the boundary. In practice, we are often interested in the injection of helicity due to ideal (or nearly ideal) motions on the boundary.

The relative helicity, $\Hr$, was motivated by applications in the solar atmosphere. Here, the magnetic field in three-dimensional space cannot generally be measured directly from telescope observations, but it is routinely observed on the Sun's photosphere (visible surface). Following \citet{1984JFM...147..133B}, these observations have been used to estimate the flux of relative helicity from the Sun's interior into its atmosphere, both on a global scale \citep{2000JGR...10510481B,2012ApJ...758...61Y}, and in individual solar active regions \citep[see][for a review]{2009AdSpR..43.1013D}. Indeed, the need to shed relative helicity into interplanetary space is understood to be the fundamental driver for coronal mass ejections \citep{1997GMS....99..119R}. In the case of numerical models and simulations, it is possible to perform the volume integral for $\Hr$ directly. Accordingly, its build-up in the solar atmosphere has been followed in a wide variety of simulations of solar active regions \citep[e.g.,][]{2000ApJ...539..944D, 2005ESASP.596E..54C, 2008PASJ...60..809M, 2011ApJ...729...97M, 2013SoPh..283..369Y, 2014SoPh..289.4453M, 2015A&A...582A..76S, 2015A&A...580A.128P, 2017A&A...601A.125P, 2018ApJ...852...82Y}. Minimisation of magnetic energy subject to conservation of $\Hr$ can also be used for the computation of force-free equilibria \citep{finin1985magnetic,1989A&A...225..156D}.

However, $\Hr$, like $h(V)$, is only a single integral over the whole domain, so cannot provide local information about the constraints on a particular magnetic field. For example, observations of the solar atmosphere suggest a highly non-uniform distribution of free energy, with the concentration of helicity in individual magnetic flux ropes believed to play an important role in their eruptivity. Moreover, recent numerical simulations of the resistive relaxation of braided magnetic flux tubes suggest the presence of additional topological constraints, over and above conservation of the global $\Hr$ \citep{2010PhRvE..81c6401D, 2011A&A...525A..57P}.

To overcome this limitation, it is desirable to compute $\Hr$ in smaller subvolumes of the overall computational domain. This was addressed by \citet{2008ApJ...674.1130L}, who divide the overall computational volume $V$ into a union of subvolumes $V=\bigcup_i D_i$. The subvolumes $D_i$ must be defined in such a way that the boundaries between them are magnetic surfaces (with vanishing normal magnetic field). In this way, the only uncontained magnetic field is on their intersection with the global boundary $\partial V$. \citet{2008ApJ...674.1130L} propose to define an ``additive self-helicity'', $H_i^{(s)}$, for each $D_i$ that is simply the relative helicity integrated only over $D_i$, with respect to a reference potential field defined locally within $D_i$. In this way, we can measure the helicity over-and-above any helicity which is forced to be present simply by the shape of the domain $D_i$. \citet{2009ApJ...702..580M} were able to relate $H_i^{(s)}$ to the kink-instability of a particular magnetic flux rope.

However, a limitation with $H_i^{(s)}$ as defined by \citet{2008ApJ...674.1130L} is the need to identify a suitable union of subvolumes $D_i$. In the absence of any further information, it seems desirable to make the decomposition very fine (i.e., many $D_i$). The finest possible decomposition is to take each $D_i$ to be an infinitesimal tube around each magnetic field line, but in that limit the local reference potential field would simply become the original magnetic field, so that $H_i^{(s)}$ (when appropriately normalised) would vanish. Therefore, in this paper, we will take a different approach that allows us to define a limiting field-line helicity for each individual magnetic field line. We will call this a relative field-line helicity because it will integrate to give $\Hr$. Note that this property is not shared by the $H_i^{(s)}$; although they add to give a relative helicity, it is relative to the sum of the local reference fields rather than to the global reference field used in $\Hr$. A further advantage of our proposed relative field-line helicity over $H_i^{(s)}$ will be its ease of computation, since it will not require computation of the potential reference field on an irregularly shaped domain.

The basic form of field-line helicity, which decomposes the original magnetic helicity $h(V)$, was given explicitly by \citet{1988A&A...201..355B} and will be defined in Section \ref{sec:prelim}. Like $h(V)$, field-line helicity has been shown to be intimately related to the topological structure of a magnetic field \citep{2013PhPl...20a2102Y, 2014JPhCS.544a2002Y,2014ApJ...787..100P}, and is a valuable tool for studying magnetic reconnection \citep{2011PhPl...18j2118Y,2015PhPl...22c2106R}. In global models of non-potential magnetic fields in the solar corona, it has recently been used successfully to identify local magnetic flux ropes \citep{2016A&A...594A..98Y,2017ApJ...846..106L}.  The main objective of the present paper is to generalise the definition of field-line helicity to give a relative field-line helicity that integrates to give $\Hr$, rather than $h(V)$. Our proposed definition is given in Section \ref{sec:def}. Numerical methods for computing relative field-line helicity are summarised in Section \ref{sec:num}, and used to explore and illustrate its behaviour for particular examples in Section \ref{sec:eg}. Although these examples by no means exhaust the possible situations where these ideas can be applied, they do illustrate some of the more typical structures that are found in solar active regions.

\section{Preliminaries} \label{sec:prelim}

Throughout this paper, we will assume our magnetic field $\Bb$ (with $\nabla\cdot\Bb=0$) to be defined in a simply-connected closed domain $V$ whose boundary $\partial V$ is a single closed surface. Our examples in Section \ref{sec:eg} will use a simple Cartesian box, which is the typical $V$ we have in mind. However, it is only the topology of $V$ (and $\partial V$) that matters for the theory in Section \ref{sec:def}.

\subsection{Relative magnetic helicity}

Importantly, we do not require $\partial V$ to be a magnetic surface, but permit non-zero $B_n := \Bb\cdot\nb$ on $\partial V$. This is the typical situation in astrophysical applications. As is well known, the magnetic helicity,
\begin{equation}
h(V) = \int_V\Ab\cdot\Bb\,\mathrm{d}^3x,
\end{equation}
is not gauge invariant in this situation, since under a gauge transformation from $\Ab$ to $\Ab'=\Ab + \nabla\chi$, the helicity becomes
\begin{equation}
h'(V) = \int_V\Ab\cdot\Bb\,\mathrm{d}^3x + \oint_{\partial V} \chi B_n\,\mathrm{d}^2x,
\label{eqn:hgauge}
\end{equation} 
so that its value depends on the gauge function $\chi$. A gauge-invariant ``relative helicity'' may be defined as
\begin{equation}
\Hr = \int_V\big(\Ab + \Ap\big)\cdot\big(\Bb - \Bp\big)\,\mathrm{d}^3x,
\label{eqn:hr}
\end{equation}
where $\Bp$ is some arbitrary reference field whose normal component $B_{{\rm p}n}$ matches $B_n$ on $\partial V$ \citep{1984JFM...147..133B,finin1985magnetic}. Typically $\Bp$ is chosen to be the unique such field that is current-free (potential) within $V$, although this is not essential. In (\ref{eqn:hr}), $\Ab$ is an arbitrary vector potential for $\Bb$ (so that $\Bb=\nabla\times\Ab$) and $\Ap$ is an arbitrary vector potential for $\Bp$ (so that $\Bp=\nabla\times\Ap$). It is easy to show from (\ref{eqn:hr}) that $\Hr$ is invariant under gauge transformation of either $\Ab$ or $\Ap$, for a given $\Bp$.

\subsection{Field-line helicity}

The concept of field-line helicity arises if we consider the magnetic helicity in a subvolume $D_i\in V$. This clearly makes sense for any material (comoving) subvolume $D_i$ provided that $B_n=0$ on $\partial D_i$, for then
\begin{equation}
h(D_i)=\int_{D_i}\Ab\cdot\Bb\,\mathrm{d}^3x
\end{equation}
would be both gauge invariant and conserved under ideal-MHD evolution. One can envisage sub-dividing $V$ into a finer and finer union of material subvolumes, in order to gain more and more detailed information about the topological structure. As mentioned in Section \ref{sec:intro}, the finest possible decomposition is where each $D_i$ is a thin magnetic flux tube surrounding a single magnetic field line, because magnetic field lines are material lines in an ideal evolution. 

\begin{figure}
\centerline{\includegraphics[width=0.6\textwidth]{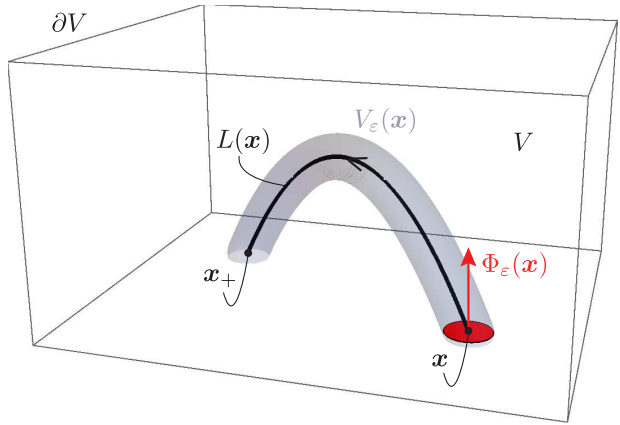}}
\caption{Definition sketch for Equation (\ref{eqn:flim}), showing a thin flux tube $V_\varepsilon(\xb)$ of radius $\varepsilon$ and magnetic flux $\Phi_\varepsilon(\xb)$ around the field line $L(\xb)$ that is rooted at $\xb\in\partial V$.}
\label{fig:tube}
\end{figure}

In this paper, however, we will consider only field lines that are rooted at both ends in the boundary $\partial V$, so that their surrounding tubular domains have $B_n\neq 0$ on the ends, violating this boundary condition. Figure \ref{fig:tube} shows such a field line,  $L(\xb)$, along with a surrounding flux tube $V_\epsilon(\xb)$ of radius $\varepsilon$ (in some cross-section). We label $L(\xb)$ by one of its endpoints $\xb\in\partial V$. Throughout this paper, $\xb$ will exclusively denote a point on the boundary $\partial V$.

Because $B_n\neq 0$ on $\partial V_\epsilon(\xb)$, we expect $h(V_\varepsilon(\xb))$ to change due to either motion of the field line endpoints or gauge transformation of $\Ab$. The latter is a more significant problem, and will be addressed in detail in this paper. First, however, we take the limit to an infinitesimally thin tube. Since $h(V_\varepsilon(\xb))$ is a volume integral, we normalise by the flux $\Phi_\varepsilon(\xb)$ of the tube and define
\begin{equation}
\mathcal{A}(\xb) = \lim_{\varepsilon\to 0}\frac{h(V_\varepsilon(\xb))}{\Phi_\varepsilon(\xb)}.
\label{eqn:flim}
\end{equation}
In the limit $\varepsilon\to 0$, the tube collapses to the line $L(\xb)$, and $\mathcal{A}(\xb)$ tends to a well-defined limit -- independent of the choice of cross-section -- which \citet{1988A&A...201..355B} calls the field-line helicity. Note that, although $\mathcal{A}(\xb)$ is defined by integrating only over the tubular volume $V_\varepsilon(\xb)$, it nevertheless contains information about how this particular flux tube interacts with the magnetic field outside, due to the fact that $\Ab$ is defined globally. In this paper, we will assume that $\Bb$ has no ergodic field lines with infinite length, as we have not shown that a well-defined limit exists in that case.

The definition (\ref{eqn:flim}) shows that $\mathcal{A}(\xb)$ has the dimensions of a magnetic flux. A more convenient formula may be obtained by writing the volume integral $h(V_\varepsilon(\xb))$ as an integral along $L(\xb)$ of cross-sectional surface integrals. In the limit $V_\varepsilon(\xb)\to L(\xb)$, the vector potential $\Ab$ is constant on each cross-section, so that
\begin{equation}
\mathcal{A}(\xb) = \lim_{\varepsilon\to 0}\frac{\int_{V_\varepsilon(\xb)}\Phi_\varepsilon(\xb)\Ab\cdot\,\mathrm{d}\lb}{\Phi_\varepsilon(\xb)} = \int_{L(\xb)}\Ab\cdot\,\mathrm{d}\lb.
\label{eqn:flh}
\end{equation}
This formula shows clearly that, were $L(\xb)$ to be a contractible closed loop, then $\mathcal{A}(\xb)$ would be gauge independent, and by Stokes' theorem it would simply be the magnetic flux linked through that loop. Clearly $\mathcal{A}(\xb)$ would then be an ideal-MHD invariant, representing the magnetic flux linked with $L(\xb)$. In our case, $L(\xb)$ is not a closed loop. Nevertheless, $\mathcal{A}(\xb)$ remains an ideal-MHD invariant provided that there are no boundary motions, and also that the gauge of $\Ab$ on the boundary is fixed. Indeed a gauge transformation from $\Ab$ to $\Ab'=\Ab+\nabla\chi$ will change the value of $\mathcal{A}(\xb)$ to
\begin{equation}
\mathcal{A}'(\xb) = \mathcal{A}(\xb) + \chi(\xb_+) - \chi(\xb),
\end{equation}
where $\xb_+$ is the other end of the field line, as in Figure \ref{fig:tube}. This is simply the limiting version of the formula (\ref{eqn:hgauge}). In fact, since $\partial V$ is a single closed surface, it follows that $\mathcal{A}(\xb)$, like $h(V)$, depends only on $\nb\times\Ab$. For if $\nb\times\Ab'=\nb\times\Ab$ on $\partial V$, then $\nb\times\nabla\chi=0$ on $\partial V$, so that $\chi$ is constant over all of $\partial V$ and $\mathcal{A}'(\xb)=\mathcal{A}(\xb)$. This would not be true for a domain where $\partial V$ is not a single connected surface, for example a spherical shell. As we will see in this paper, the art of working with field-line helicity is to choose an informative gauge for $\nb\times\Ab$.

Finally, we note the relation between $\mathcal{A}(\xb)$ and the overall helicity $h(V)$. When the field lines $L(\xb)$ partition the whole volume (i.e., there are no closed or ergodic field lines), integrating (\ref{eqn:flim}) over all field lines, weighted by magnetic flux, will give $h(V)$ \citep{1988A&A...201..355B}. In other words,
\begin{equation}
\frac{1}{2}\oint_{\partial V}\mathcal{A}|B_n|\,\mathrm{d}^2x = h(V),
\label{eqn:htota}
\end{equation}
where the factor half arises because each field line has two end-points on $\partial V$. 

\section{Definition of relative field-line helicity} \label{sec:def}

The aim of this paper is to generalise the formula (\ref{eqn:htota}) to the relative helicity $\Hr$. In other words, we would like to define a ``relative field-line helicity'' $\Ar(\xb)$ for each field line that is invariant under an ideal-MHD evolution and satisfies
\begin{equation}
\frac{1}{2}\oint_{\partial V}\Ar|B_n|\,\mathrm{d}^2x = \Hr,
\label{eqn:hrfromflh}
\end{equation}
for a fixed choice of reference field. We will firstly show that there are many different ways to define such an $\Ar$, and will go on to make some specific choices that we believe are physically reasonable.

\subsection{Towards a definition}

One option is to use $\mathcal{A}(\xb)$ directly, but restrict the gauge $\Ab$. We know that, whatever the gauge, $\mathcal{A}(\xb)$ is invariant under an ideal-MHD evolution for every field line provided that $\nb\times\Ab$ on $\partial V$ remains fixed in time. But in an arbitrary gauge, condition (\ref{eqn:hrfromflh}) will not be satisfied, since $h(V)\neq \Hr$ in general. Nevertheless, there are a restricted family of gauges of $\Ab$ where $h(V)=\Hr$. To see this, write (\ref{eqn:hr}) in the form
\begin{equation}
\Hr = \int_V\Ab\cdot\Bb\,\mathrm{d}^3x + \oint_{\partial V}\Ab\times\Ap\cdot\nb\,\mathrm{d}^2x - \int_{V}\Ap\cdot\Bp\,\mathrm{d}^3x.
\label{eqn:finn}
\end{equation}
It is well known that the boundary term vanishes if we restrict $\nb\times\Ab=\nb\times\Ap$ on $\partial V$ \citep{1988PhFl...31.2214B,1988A&A...201..355B}. This is always possible since $B_{{\rm p}n}=B_n$ on $\partial V$ (Section \ref{sec:match} shows how to construct such an $\Ab$, given an arbitrary $\Ap$). We then need only choose $\Ap$ such that $\int_V\Ap\cdot\Bp\,\mathrm{d}^3x=0$, and we will have $\Hr=h(V)$ and hence (\ref{eqn:hrfromflh}). In general, this is not possible, since $\Bp$ may have a closed-loop field line $L$ within $V$ whose gauge-invariant $\oint_L\Ap\cdot\mathrm{d}\lb$ does not vanish. However, in the case where $\Bp$ is a potential field, it cannot have any closed-loop field lines, and we can always find a gauge for $\Ap$ such that $\int_V\Ap\cdot\Bp\,\mathrm{d}^3x=0$ \citep[e.g.,][]{1988PhFl...31.2214B,1988A&A...201..355B}.

Even when $\Bp$ is a potential field, the conditions that  $\int_V\Ap\cdot\Bp\,\mathrm{d}^3x=0$ and $\nb\times\Ab=\nb\times\Ap$ on $\partial V$ do not uniquely define $\mathcal{A}(\xb)$, because there is still freedom in the choice of $\Ap$. For example, if we have such a gauge $\Ap$ and change to $\Ap' =\Ap+\nabla\phi$, then
\begin{equation}
\int_V\Ap'\cdot\Bp\,\mathrm{d}^3x = \oint_{\partial V}\phi B_n\,\mathrm{d}^2x.
\end{equation}
Provided $B_n\neq 0$, this integral may still vanish even if $\phi\neq 0$ on $\partial V$.

\subsection{General definition}

In practice, it may be inconvenient or (when $\Bp$ is non-potential) impossible to find a gauge $\Ap$ such that $\int_V\Ap\cdot\Bp\,\mathrm{d}^3x=0$. Therefore, we propose instead an alternative definition for relative field-line helicity that does not impose this requirement on $\Ap$. Namely,
\begin{equation}
\Ar(\xb) := \mathcal{A}(\xb) - \mathcal{A}_{\rm p}(\xb),
\label{eqn:ar}
\end{equation}
where $\mathcal{A}(\xb)$ is the usual field-line helicity (\ref{eqn:flh}) and 
\begin{equation}
\mathcal{A}_{\rm p}(\xb) = \int_{L_{\rm p}(\xb)}\Ap\cdot\,\mathrm{d}\lb,
\end{equation}
which is the field-line helicity of the reference field $\Bp$ on its own field line rooted at the same point $\xb\in \partial V$. Importantly, our definition assumes the gauge condition on $\Ab$ that
\begin{equation}
\nb\times\Ab=\nb\times\Ap \quad \textrm{on $\partial V$}.
\end{equation}
It follows that $\Ar(\xb)$ will satisfy (\ref{eqn:hrfromflh}), whatever the gauge of $\Ap$. Note that, unlike $\mathcal{A}(\xb)$, this $\Ar(\xb)$ will not, in general, have the same value at both ends of a given field line $L(\xb)$, since the corresponding reference field line $L_{\rm p}(\xb)$ at each end will be different.

A nice property of $\Ar(\xb)$ is that $\Ar(\xb)=0$ whenever $\Bp=\Bb$ all along $L(\xb)$ for some $\xb\in\partial V$. For in that case, $L(\xb)=L_{\rm p}(\xb)$ and we must have $\Ab=\Ap+\nabla\chi$ along this line. Then
\begin{equation}
\Ar(\xb) = \int_{L(\xb)}(\Ab - \Ap)\cdot\,\mathrm{d}\lb = \int_{L({\xb})}\nabla\chi\cdot\,\mathrm{d}\lb.
\end{equation}
However, since $\partial V$ is a connected surface, and $\nb\times\nabla\chi=0$ on this surface thanks to our gauge restriction, it follows that $\chi$ is constant on $\partial V$ and hence $\Ar(\xb)=0$. This property makes $\Ar$ useful for identifying non-potential regions within $\Bb$, and will not generally be shared by the basic field-line helicity $\mathcal{A}$, even in a gauge where $\int_V\Ap\cdot\Bp\,\mathrm{d}^3x=0$ overall.

\subsection{Gauge dependence and physical meaning}

Unfortunately, our definition of $\Ar$ in (\ref{eqn:ar}), even with the restriction that $\nb\times\Ab=\nb\times\Ap$ on $\partial V$, is still not uniquely defined, and depends on the gauge of $\Ap$. To see this, consider a gauge transformation from $\Ap$ to $\Ap'=\Ap + \nabla\phi$. To preserve the boundary restriction, we must also change the gauge of $\Ab$ to $\Ab'=\Ab + \nabla\chi$, where $\nb\times\nabla(\phi-\chi)=0$ on $\partial V$. It follows that the two gauge functions can differ on $\partial V$ only by a global constant (since $\partial V$ is connected). Using this fact, we find that $\Ar(\xb)$ goes to
\begin{eqnarray}
{\Ar}'(\xb) &=& \Ar(\xb) + \int_{L(\xb)}\nabla\chi\cdot\,\mathrm{d}\lb - \int_{L_{\rm p}(\xb)}\nabla\phi\cdot\,\mathrm{d}\lb,\\
&=& \Ar(\xb) + \phi(\xb_+) - \phi(\xb_{{\rm p}+}).
\label{eqn:argauge}
\end{eqnarray}
When $L(\xb)\neq L_{\rm p}(\xb)$, the other ends of the field lines (denoted $\xb_+$ and $\xb_{{\rm p}+}$) will differ, so the value of $\Ar(\xb)$ will change.

\begin{figure}
\centerline{\includegraphics[width=\textwidth]{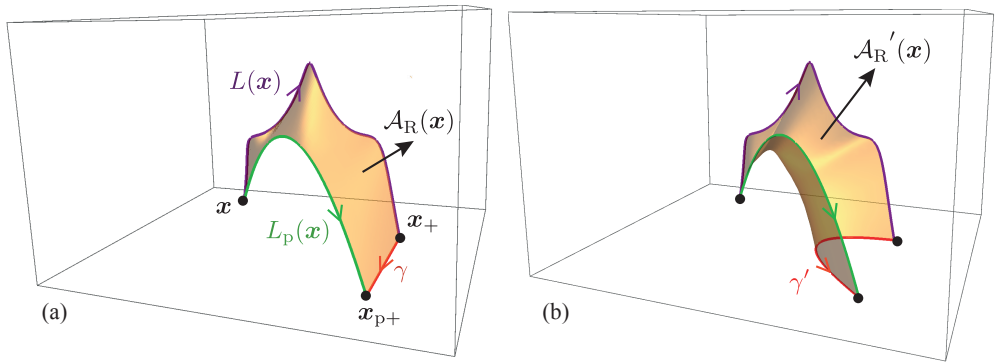}}
\caption{Interpretation of $\Ar(\xb)$ as a magnetic flux, for two different gauges in (a) and (b). The field lines $L(\xb)$ and $L_{\rm p}(\xb)$ are the same in each case, but the possible curves $\gamma$ and $\gamma'$ that complete the loop by linking $\xb_+$ and $\xb_{{\rm p}+}$ on $\partial V$ differ between the two gauges, so that $\Ar(\xb)\neq{\Ar}'(\xb)$.}
\label{fig:flux}
\end{figure}

To see the physical meaning of this gauge dependence, we can interpret $\Ar(\xb)$ as a magnetic flux \citep[cf.][]{2016A&A...594A..98Y}. This is illustrated in Figure \ref{fig:flux}, where $L(\xb)$ is a field line of $\Bb$ and $L_{\rm p}(\xb)$ is a field line of $\Bp$ rooted at the same point $\xb$. The curve $\gamma\subset\partial V$ in Figure \ref{fig:flux}(a) closes the loop and defines a surface in $V$ bounded by $L(\xb)$, $\gamma$, and $-L_{\rm p}(\xb)$. If $\gamma$ is chosen so that $\int_\gamma\Ap\cdot\,\mathrm{d}\lb=0$, then $\Ar(\xb)$ will be precisely the magnetic flux through this surface, by Stokes' theorem. Provided the field-line endpoints and $\nb\times\Ap$ on $\partial V$ and remain fixed in time, this flux will be an ideal-MHD invariant. In any gauge, it is possible to find such a curve $\gamma$ along which the line integral vanishes, as proven by \citet{2016A&A...594A..98Y}. In fact, there are an infinite number of such curves (whose corresponding surfaces all have the same flux). The possible $\gamma$ depend on the gauge of $\nb\times\Ap$, as sketched in Figure \ref{fig:flux}(b) which shows a different curve $\gamma'$ arising in a different gauge $\Ab'$. This loop encloses a different amount of magnetic flux, corresponding to the different value of ${\Ar}'(\xb)$ in this new gauge. In summary, the gauge dependence of $\Ar$ may be viewed simply as a choice of how to close  flux surfaces on the boundary. Whichever choice is made, $\Ar(\xb)$ is an ideal-MHD invariant.

At first sight, it may seem from (\ref{eqn:argauge}) that we can choose any distribution of $\Ar$ that we like on field lines that intersect $\partial V$, simply by changing the gauge $\phi$. Indeed, this is true. However, choosing $\phi$ in this way \emph{requires you to first know $\Ar$ for $\Bb$ in some initial gauge}. If we choose the gauge \emph{a priori}, based only on $\Bp$, then the resulting $\Ar$ can give us meaningful information about $\Bb$.

In Section \ref{sec:eg}, we will illustrate the effect on $\Ar$ of some different choices of $\Ap$ in particular examples. However, we will first propose a general candidate for the ``best'' choice of gauge, that can be used to uniquely define $\Ar$.

\subsection{Minimal gauge}

Since $\Ar$ depends only on the distribution of $\nb\times\Ap$ on $\partial V$, our idea is to choose this gauge by minimising the integral $\oint_{\partial V}|\nb\times\Ap|^2\,\mathrm{d}^2x$ over the whole boundary. Intuitively, this will give us the smallest overall boundary contribution to $\Ar$, for a given reference field $\Bp$ (whether potential or not). In Appendix \ref{app:var}, we prove that this integral is minimised if 
\begin{equation}
\nabla_h\cdot\Ap=0 \quad \textrm{on $\partial V$},
\label{eqn:astar}
\end{equation}
where $\nabla_h\cdot$ denotes the two-dimensional divergence in the plane of $\partial V$. This is the so-called ``universal gauge'' condition of \citet{2006astro.ph..6694H}. Although this condition does not specify $\Ap$ uniquely within $V$, it does specify $\nb\times\Ap$ uniquely on $\partial V$, and hence uniquely specifies $\Ar$. To see this, suppose both $\Ap$ and $\Ap'=\Ap+\nabla\phi$ satisfy (\ref{eqn:astar}). Then we must have $\nabla_h^2\phi=0$, but since $\partial V$ is a closed surface, this implies that $\phi$ is constant on $\partial V$, so that $\nb\times\Ap'=\nb\times\Ap$. Note that $\mathcal{A}_{\rm p}$ does not necessarily vanish in this gauge when $\Bp$ is a potential field -- we will see this in Section \ref{sec:eg}.

In the rest of this paper, we will denote vector potentials satisfying (\ref{eqn:astar}) by $\Ab^*$ or $\Ap^*$ , the corresponding field-line helicities by $\mathcal{A}^*$ and $\mathcal{A}_{\rm p}^*$, and the corresponding relative field-line helicity by $\Ar^*$. In Section \ref{sec:eg}, we will compute $\Ar^*$ for some example magnetic fields, and compare it to $\Ar$ in some other gauges.

\subsection{Summary of the proposed definition} \label{sec:sum}

To summarise, let $\Ap^*$ be a vector potential for $\Bp$ satisfying
\begin{equation}
\nabla_h\cdot\Ap^*=0 \quad \textrm{on $\partial V$},
\end{equation}
and let $\Ab^*$ be a vector potential for $\Bb$ satisfying
\begin{equation}
\nb\times\Ab^*=\nb\times\Ap^* \quad \textrm{on $\partial V$}.
\end{equation}
Then our proposed definition of relative field line helicity at $\xb\in\partial V$ is the difference
\begin{equation}
\Ar^*(\xb) := \mathcal{A}^*(\xb) - \mathcal{A}^*_{\rm p}(\xb),
\end{equation}
where $\mathcal{A}^*$, $\mathcal{A}^*_{\rm p}$ are the field-line helicities of $\Bb$ and $\Bp$ in these two gauges, defined along their respective magnetic field lines.

\section{Numerical methods} \label{sec:num}

The numerical code used for this paper was written in Python, with a supporting Fortran module for fast tracing of the magnetic field lines (taking advantage of OpenMP if available). This code is open source and available at \url{https://github.com/antyeates1983/flhtools}. Although the definitions in Section \ref{sec:def} apply to more general domains, this code is specific to a Cartesian domain.

For field-line tracing, the code uses the second-order midpoint method with adaptive step-size. The computations in this paper used simple linear interpolation of $\Ab$ and $\Bb$ to trace field lines and compute $\mathcal{A}$, demonstrating that sophisticated numerical methods are not required to work with field-line helicity. However, to facilitate computation of $\Bp$, $\Ab$ and $\Ap$, the code uses a staggered mesh where each component of $\Bb$ (or $\Bp$) is located at the centre of the corresponding cell face and the components of $\Ab$ (or $\Ap$) are located on the corresponding cell edges \citep{1966ITAP...14..302Y}.

\subsection{Reference potential field} \label{sec:pf}

In a Cartesian box, the potential reference field $\Bp$ that matches $B_n$ on all six boundary faces is straightforward to compute. We write $\Bp=\nabla u$ leading to the three-dimensional Laplace equation $\nabla^2 u = 0$. On the staggered mesh, the potential $u$ is located at the centre of each three-dimensional cell, so that each component of $\Bp$ may be computed from $u$ by a central difference. We approximate the Laplacian operator $\nabla^2$ with a second-order (7-point) finite-difference stencil, for which the solution is efficiently and simply obtained using a fast-Poisson technique with Neumann boundary conditions \citep{1992nrfa.book.....P}. With this technique, the current density $\nabla\times\Bp$ vanishes to machine precision when computed from $\Bp$ with central differences.

\subsection{Minimal-gauge vector potential} \label{sec:astar}

To find $\Ab^*$ (or $\Ap^*$), we have developed a numerical routine that starts from $\Ab$ (or $\Ap$) in an arbitrary gauge, and performs the necessary gauge transformation. In particular, for the case of $\Ab$, we seek a scalar function $\chi(x,y,z)$ on $V$ such that $\Ab^*=\Ab + \nabla\chi$. The condition $\nabla_h\cdot\Ab^*=0$ on $\partial V$ requires that $\overline{\chi}:=\chi|_{\partial V}$ satisfy the two-dimensional Poisson equation 
\begin{equation}
\nabla_h^2\overline{\chi} = -\nabla_h\cdot\Ab
\label{eqn:poisson}
\end{equation}
over this whole boundary $\partial V$. Since $\partial V$ is a closed surface, the right-hand side necessarily satisfies the compatibility condition $\oint_{\partial V}\nabla_h\cdot\Ab\,\mathrm{d}^2x=0$, so that the solution for $\overline{\chi}$ exists and is unique up to an additive constant $\overline{\chi}_0$.

Equation (\ref{eqn:poisson}) is solved numerically by a finite-difference method, approximating the Laplacian operator with the standard 5-point stencil. For our Cartesian domain, care must be taken to couple values of $\overline{\chi}$ on neighbouring faces so that the normal component of $\nabla_h\overline{\chi}$ is continuous. (Strictly speaking,  we are approximating a weak solution to (\ref{eqn:poisson}), since the differential operators in (\ref{eqn:poisson}) are not defined on the edges of the cube.) This coupling between faces precludes the use of a fast-Poisson solver, but a direct solution of the resulting (sparse) linear system remains practical for this two-dimensional problem.  Although $\Ab^*$ does not depend on the additive constant $\overline{\chi}_0$, we ensure that the linear system is invertible by fixing $\overline{\chi}=0$ at one vertex of $\partial V$.

Given the solution for $\overline{\chi}$, we can then choose any arbitrary extension of $\chi$ into the interior of $V$, since this choice does not affect $\mathcal{A}^*$. It is simplest to take $\nabla^2\chi=0$ in the volume, and solve this three-dimensional Laplace equation with Dirichlet boundary conditions $\chi=\overline{\chi}$ on $\partial V$. This can be solved using a standard fast-Poisson method.

\subsection{Matching the reference gauge} \label{sec:match}

Although not required for computing $\Ar^*$, we have also implemented a numerical routine for changing the gauge of $\Ab$ to $\Ab'$ so that $\nb\times\Ab'=\nb\times\Ap$ on $\partial V$, given some $\Ap$. This will be used in Section \ref{sec:eg} to illustrate the computation of $\Ar$ in gauges other than the minimal gauge.

Writing $\Ab' = \Ab + \nabla\chi$, we need to compute $\chi(x,y,z)\in V$ such that $\nb\times\Ab' = \nb\times\Ap$ on $\partial V$. This may be done by first solving for $\chi|_{\partial V}=\overline{\chi}$ on $\partial V$, then extending to a solution on the interior of $V$. We need $\overline{\chi}$ to satisfy $\nabla_h\overline{\chi} = \Ab_{\mathrm{p}h} - \Ab_h$, so take the divergence and solve the resulting two-dimensional Poisson equation
\begin{equation}
\nabla_h^2\overline{\chi} = \nabla_h\cdot(\Ap - \Ab)
\label{eqn:chimatch}
\end{equation}
on $\partial V$. As in Section \ref{sec:astar}, a solution exists because the compatibility condition is satisfied. Since the Neumann boundary condition $\nb\cdot\nabla\overline{\chi}=\nb\cdot(\Ap - \Ab)$ is fixed on the edges of each face, it is possible to solve (\ref{eqn:chimatch}) separately on each face. This means that a fast-Poisson solver can readily be used, unlike in Section \ref{sec:astar}. This gives $\overline{\chi}$ on each face $S_i$ up to a constant $\overline{\chi}_i$, for $i=1,\ldots,6$. One of these constants may be chosen arbitrarily, and the other five are then easily determined by imposing continuity of $\overline{\chi}$ at each edge. (Although there are 12 edges, the corresponding jumps in $\overline{\chi}$ are not independent, so the problem is well-posed.)

Once the continuous solution for $\overline{\chi}$ is obtained, this may be extended to $\chi$ in the interior of $V$ in the same way as Section \ref{sec:astar}, namely solving the three-dimensional Laplace equation $\nabla^2\chi=0$ in $V$ with Dirichlet boundary conditions $\chi=\overline{\chi}$ on $\partial V$. Again, this choice of interior distribution has no effect on the field-line helicity.

\section{Examples} \label{sec:eg}

To investigate the behaviour of $\Ar$, we consider two well-known solar active region models from the literature. These exhibit important magnetic structures found in MHD simulations, namely sheared magnetic arcades and magnetic flux ropes. They are chosen  specifically to enable validation against the results of \citet{2016SSRv..201..147V}, who made a careful study of the numerical computation of $\Hr$ in these particular examples (among others).

\subsection{Low-Lou equilibrium}

First, we consider one of the class of nonlinear force-free equilibria derived by \citet{1990ApJ...352..343L}, which is given analytically except for the solution of a single ordinary differential equation. Our calculations start from a three-dimensional datacube of $\Bb$, computed by \citet{2016SSRv..201..147V} and kindly shared by these authors. This corresponds to the specific solution with parameters $n=1$, $l=0.3$, $\phi=\pi/4$, in a Cartesian box $V=[-1,1]\times[-1,1]\times[0,2]$. We used four datacubes with uniform mesh spacing $\Delta x = 2/32$, $2/64$, $2/128$, and $2/256$, respectively. As described in Section \ref{sec:num}, our numerical code uses a staggered mesh, so we first interpolated the original $\Bb$ components from cell vertices to the centres of the cell faces, using trilinear interpolation. We also tested tricubic interpolation here but found it unnecessary.

\begin{figure}
\centerline{\includegraphics[width=\textwidth]{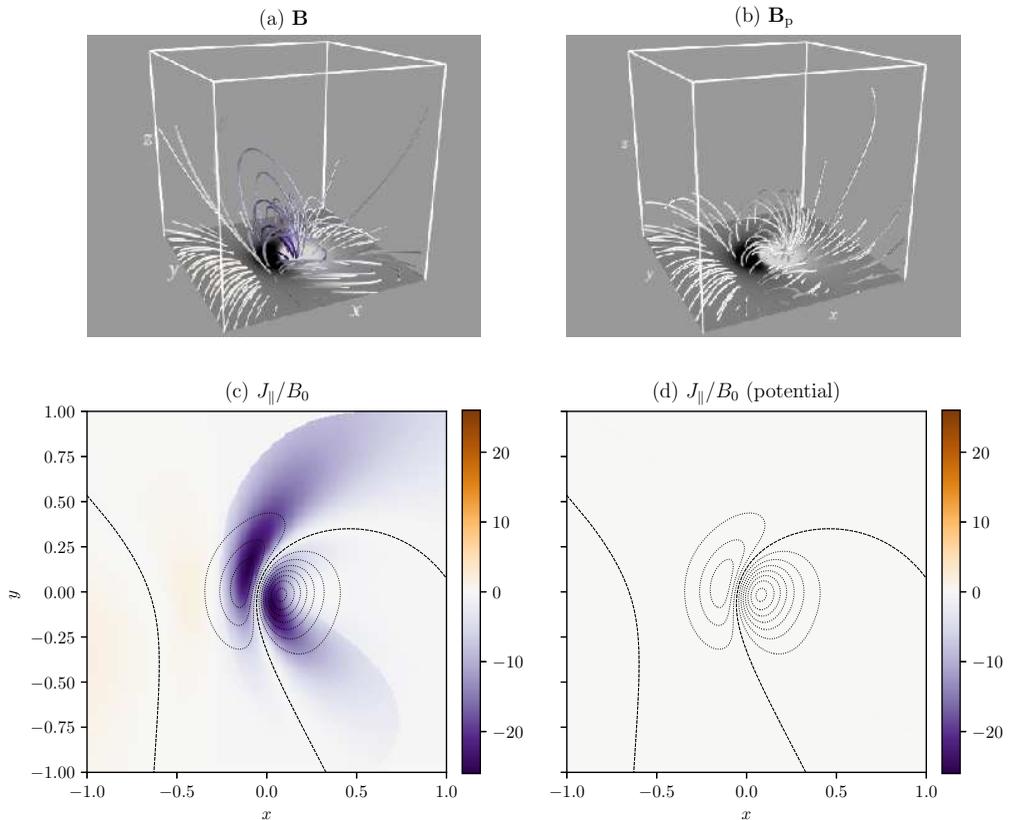}}
\caption{Field-line structure of the Low-Lou example $\Bb$ (a) and its corresponding potential reference field $\Bp$ (b), for the datacube with $\Delta x = 2/128$. The field lines are coloured purple-orange according to their integrated parallel current, shown in the $z=0$ plane in panels (c) and (d). On the lower boundary, the normal field $B_z(x,y,0)=B_{{\rm p}z}(x,y,0)$ is shown by greyscale contours (a and b) or dashed lines (c and d).}
\label{fig:llfl}
\end{figure}

Figure \ref{fig:llfl} shows the magnetic field lines for this magnetic field $\Bb$ (panel a) and for the corresponding reference potential field $\Bp$ (panel b). The latter was computed numerically to match $B_n$ on all six boundaries, using the method described in Section \ref{sec:pf}. In Figure \ref{fig:llfl}, each field line $L({\xb})$ is coloured by its integrated parallel current density, defined as
\begin{equation}
J_\parallel(\xb) = \int_{L({\xb})}\nabla\times\Bb\cdot\,\mathrm{d}\lb
\end{equation}
and normalised by $B_0$, the mean value of $|B_z(x,y,0)|$. Maps of $J_\parallel/B_0$ for all field lines seeded from the $z=0$ boundary are shown in panels (c) and (d).
In the Low-Lou equilibrium, the electric currents are smoothly distributed within the arcade of magnetic loops surrounding the central polarity-inversion line. For $\Bp$, we have $\nabla\times\Bp\equiv\boldsymbol{0}$ and hence $J_\parallel$ vanishes (to machine precision).

Next, we have computed maps of the field-line helicities $\mathcal{A}$, $\mathcal{A}_{\rm p}$ and the relative field-line helicity $\Ar:=\mathcal{A} - \mathcal{A}_{\rm p}$  using four different gauges. These include the minimal gauge $\Ap^*$, but also three others, denoted $\Ap^a$, $\Ap^b$, $\Ap^c$ and defined as follows. All three satisfy the \citet{2000ApJ...539..944D} condition that $A_z=0$. The first is integrated upward in from the lower $z$ boundary,
\begin{equation}
\Ap^a(x,y,z) = {\Ab}_0^a(x,y)  - \hat{\zb}\times \int_0^z \Bb(x,y,z')\,\mathrm{d}z',
\end{equation}
and uses the Coulomb-gauge form $\Ab_0^a=\nabla\times(\Psi\hat{\zb})$ for the boundary term, with $\Psi$ computed by solving the Poisson equation $\nabla_h^2\Psi=-B_z(x,y,0)$ (we separate the monopole component and solve for the remainder using periodic boundary conditions). The second gauge is similar but integrated down from the upper $z$ boundary, so that
\begin{equation}
\Ap^b(x,y,z) = {\Ab}_0^b(x,y) + \hat{\zb}\times\int_z^2 \Bb(x,y,z')\,\mathrm{d}z'
\end{equation}
and $\Ab_0^b$ is now computed on $z=2$ from $B_z(x,y,2)$.
The third is an alternative DeVore gauge that still has the form
\begin{equation}
\Ap^c = {\Ab}_0^c(x,y) - \hat{\zb}\times \int_0^z \Bb(x,y,z')\,\mathrm{d}z',
\end{equation}
but uses the alternative boundary term
\begin{equation}
\Ab_0^c=-\frac{1}{2}\left(\int_{-1}^y B_z(x,y',0)\,\mathrm{d}y'\right)\hat{\xb} + \frac{1}{2}\left(\int_{-1}^x B_z(x',y,0)\,\mathrm{d}x'\right)\hat{\yb}.
\end{equation}
There is nothing special about these chosen gauges, other than their simplicity of computation. The latter has led to the use of $\Ap^a$ and $\Ap^b$ for computation of $\Hr$ \citep[e.g.][]{2016SSRv..201..147V}, where, of course, the choice of gauge does not affect the result, except for numerical error.

\begin{figure}
\centerline{\includegraphics[width=\textwidth]{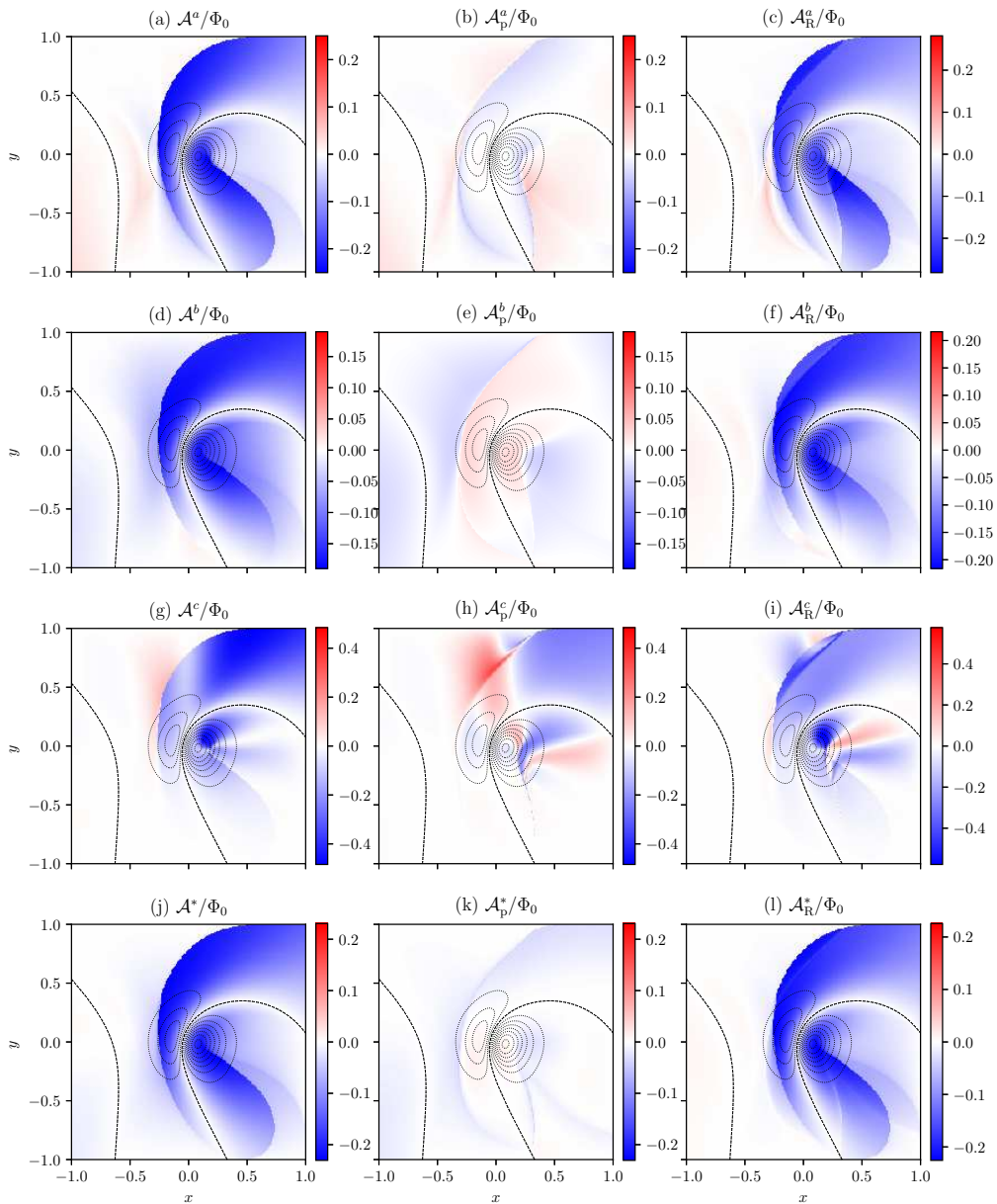}}
\caption{Field-line helicity on $z=0$ for the Low-Lou example in four different gauges (top to bottom). The left and middle panels show $\mathcal{A}$ for the original and potential fields, respectively, while the right panel shows their difference $\Ar$ in each case, which is the relative field-line helicity. The definitions of the vector potentials are given in the text, but note that panels (j) to (l) show the minimal gauge $\Ab^*$. Dashed lines show contours of $B_z=B_{{\rm p}z}$ for context. All of the field-line helicities are normalised by the boundary flux $\Phi_0=\frac{1}{2}\int_{z=0}|B_n|\,\mathrm{d}^2x$.}
\label{fig:lla}
\end{figure}

Figure \ref{fig:lla} shows maps of $\mathcal{A}$, $\mathcal{A}_{\rm p}$, and $\Ar$ for each of the four gauges, computed at resolution $256\times 256$ from the datacube with resolution $\Delta x = 2/128$. These maps show only the $z=0$ boundary. For the first three gauges, $\nb\times\Ab$ has been matched to $\nb\times\Ap$ using the gauge-matching procedure described in Section \ref{sec:num}. For $\Ap^*$, this is not necessary as both $\Ab$ and $\Ap$ are independently matched to the minimal gauge using the procedure in Section \ref{sec:astar}.

The first important observation is that all three distributions depend on the gauge choice $\nb\times\Ap$. In fact, the results from $\Ap^a$ and especially $\Ap^b$ are quite close to $\Ap^*$ in this case. This is because, for this specific $\Bb$, the majority of $|B_n|$ is located on the lower boundary, where $\Ab_0^a$ and $\Ab_0^b$ satisfy the solenoidal condition $\nabla_h\cdot\Ab=0$. The gauge $\Ap^c$ (Figures \ref{fig:lla}g--i) lacks this property, and gives more radically different results from $\Ap^*$. Notice that $\Ap^c$ leads to larger maximum values of $|\Ar|$, and that these are more localised than either the regions of strongest $|\Ar^*|$ or of strongest $J_\parallel$ (Figure \ref{fig:llfl}c).

Notice that the field-line helicity of the reference field, $\mathcal{A}_{\rm p}$, is weaker than $\mathcal{A}$ in both $\Ab^a$ and $\Ab^b$, and even more so in $\Ab^*$. This is a good indication that these are ``sensible'' gauge choices, since $\Bp$ is the minimum-energy field and cannot have significantly twisted field lines \citep[it can, however, still exhibit topological structure due to the non-uniform boundary conditions $B_{{\rm p}n}$, cf.][]{2014ApJ...787..100P}. Similarly, although $\mathcal{A}^*$ is primarily located within the main current-carrying region (cf. Figure \ref{fig:llfl}c), it does not completely vanish outside of this region. The boundary of this current-carrying region appears as a sharp separatrix between field lines with both ends on the lower boundary $z=0$ and field lines with only one end on this boundary. Notice that the location of this separatrix differs between $\Bb$ and $\Bp$, so that the relative field-line helicity map $\Ar$ inherits discontinuous jumps from both $\mathcal{A}$ and $\mathcal{A}_{\rm p}$.

\begin{figure}
\centerline{\includegraphics[width=0.8\textwidth]{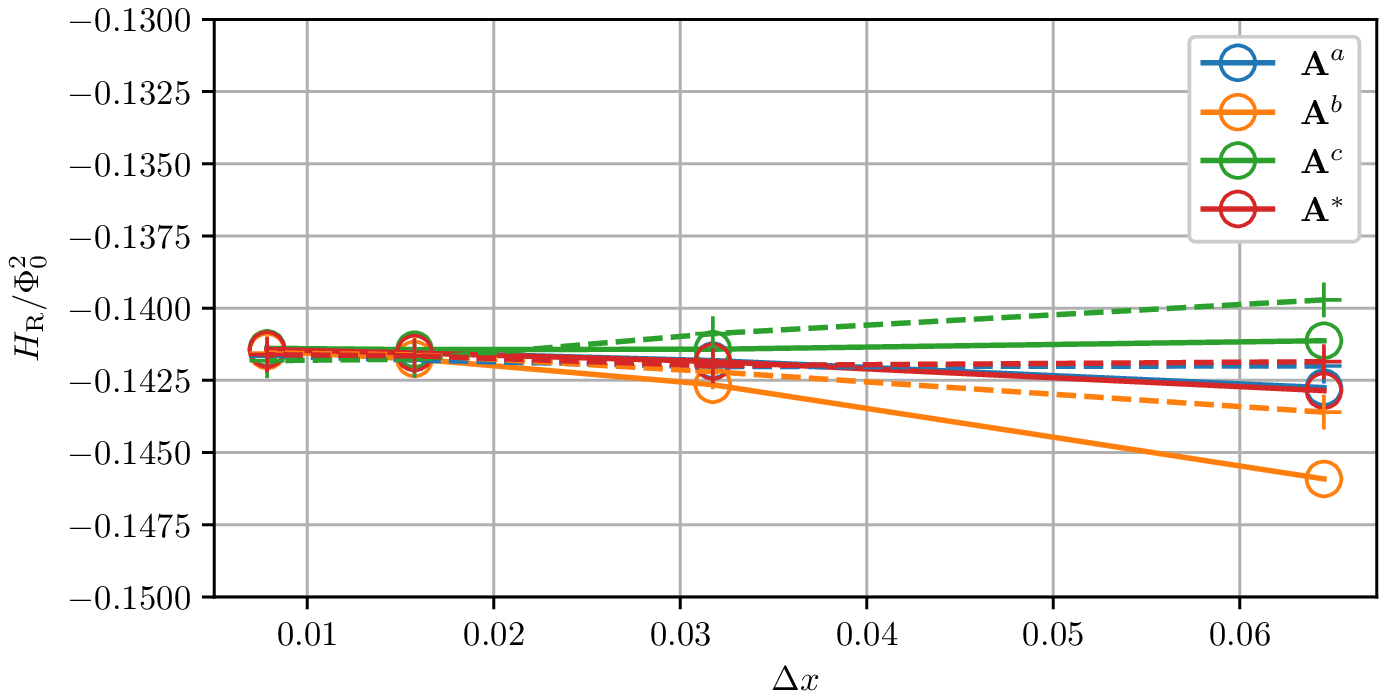}}
\caption{Numerically estimated relative helicity $\Hr$ for the Low-Lou example, as a function of the numerical mesh spacing, using four different gauges as shown in the legend. In each case, solid lines/circles show direct volume-integrated $\Hr$ using the composite trapezium rule, while dashed lines/pluses are computed from by integrating $\Ar|B_n|/2$  over all six boundary faces. The value of $\Hr$ is normalised by the square of the boundary flux $\Phi_0$.}
\label{fig:llh}
\end{figure}

As a validation exercise, we have  verified that the all of the $\Ar$ in Figure \ref{fig:lla} integrate to give the same $\Hr$, as they should by definition. Figure \ref{fig:llh} shows the resulting $\Hr$ for each of the four gauges, as a function of the  mesh spacing $\Delta x$ for the four datacubes. Solid curves show volume integrals of $(\Ab + \Ap)\cdot(\Bb - \Bp)$, using the composite trapezium rule, whereas dashed curves show estimates of $\Hr$ from integrating $\Ar$ using equation (\ref{eqn:hrfromflh}). For the latter computations, field lines were traced from a mesh with spacing $\Delta x \times \Delta x$ on each of the six boundary faces, and the integral then estimated with the composite trapezium rule. It is clear from Figure \ref{fig:llh} that taking a finer mesh leads to convergence to a common value of $\Hr$ between the different gauges, whether computed by volume integration or from $\Ar$. Moreover, this common value is consistent with the computations of \citet[][see their Figure 7a]{2016SSRv..201..147V}. 

\subsection{Titov-D\'emoulin equilibria}

\begin{figure}
\centerline{\includegraphics[width=\textwidth]{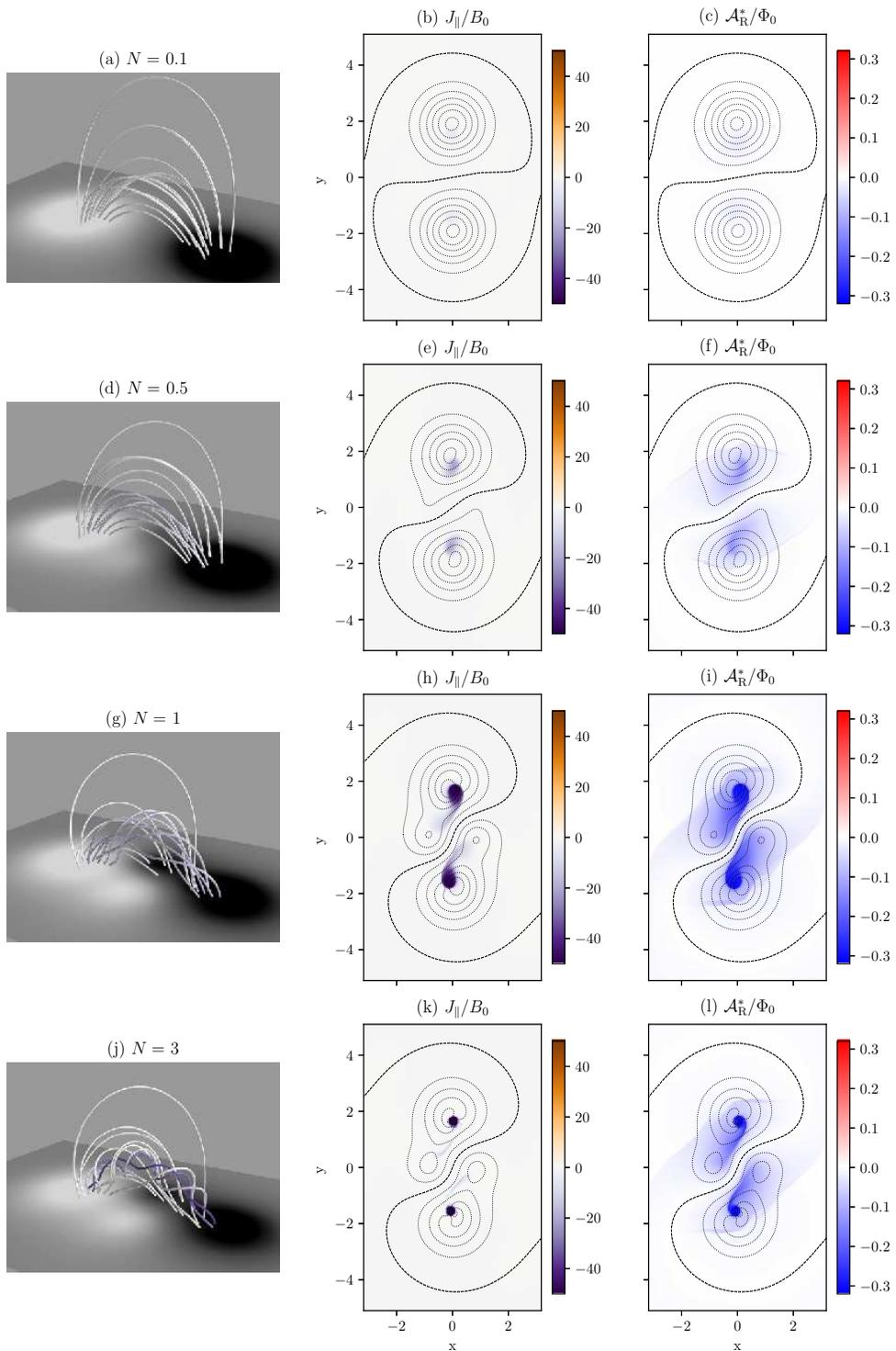}}
\caption{The Titov-D\'emoulin solution for increasing (top to bottom) values of the twist parameter $N$, showing the magnetic field lines, field-line integrated parallel current $J_\parallel/B_0$, and relative field-line helicity $\Ar^*$ in each case. The field lines are coloured by $J_\parallel$, with the same colour scale as the middle column. The colour scales are fixed; actual maxima of $|J_\parallel/B_0|$ are $4.6$, $21.2$, $53.3$, $199.4$, and of $|\Ar^*/\Phi_0|$ are $0.03$, $0.13$, $0.31$, $0.32$, respectively.}
\label{fig:tdw}
\end{figure}

More realistic models of non-potential solar active regions are given by the Titov-D\'emoulin family of equilibria \citep{1999A&A...351..707T}, in which the free magnetic energy is localised to a toroidal current channel contained within a surrounding potential field. We utilise datacubes of four such equilibria computed and kindly shared by \citet{2016SSRv..201..147V}. For details of the specific construction of these equilibria, see that paper. Here we just note that the domain is the Cartesian volume $V = [-3.18, 3.18]\times[-5.10, 5.10]\times[0, 4.56]$, and that the primary difference between the four cases is the twist of magnetic field lines in the current channel. The cases are denoted $N=0.1, 0.5, 1, 3$, where $N$ is the (approximate) number of twists made by a field line along the current channel. Corresponding magnetic field line plots are shown in Figure \ref{fig:tdw} (a, d, g, j). Note that the boundary conditions $B_n$ and potential magnetic energy also differ between the cases. The relative free energies (ratio of non-potential to total magnetic energy) are $\approx 0.002, 0.03, 0.20,$ and $0.15$, respectively. In all computations here, we use datacubes with $\Delta x = 0.06$.

\begin{figure}
\centerline{\includegraphics[width=\textwidth]{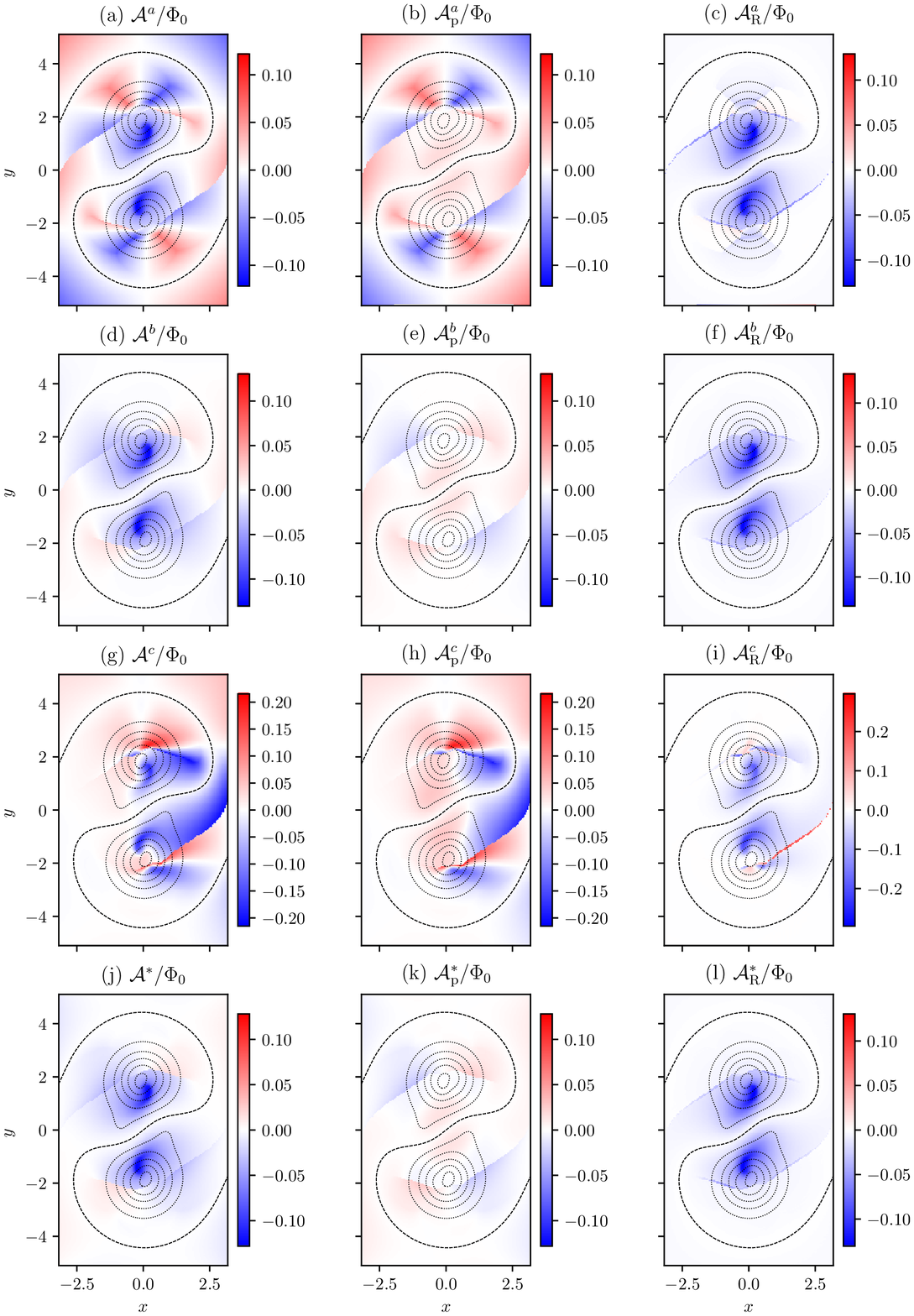}}
\caption{Field-line helicity on $z=0$ for the $N=0.5$ Titov-D\'emoulin model in four different gauges (top to bottom). The format is the same as Figure \ref{fig:lla}.}
\label{fig:tda}
\end{figure}

We begin our analysis with the same computation of $\mathcal{A}$, $\mathcal{A}_{\rm p}$, and $\Ar$ in four different gauges as for the Low-Lou equilibrium. These are shown in Figure \ref{fig:tda} for one of the datacubes: $N=0.5$. Once again, the results are clearly dependent on the gauge. However, in this case, the downward DeVore-Coulomb gauge ($\Ap^b$) gives very similar results to the minimal gauge ($\Ap^*$), whereas the upward DeVore-Coulomb gauge ($\Ap^a$) gives rather different results, along with $\Ap^c$. In all four cases, $\mathcal{A}$ agrees well with $\mathcal{A}_{\rm p}$ outside of the toroidal current channel (which appears as two tear-drop shaped footpoints on this boundary). However, for $\Ap^b$ and $\Ap^*$, the current channel has a much higher strength of $|\mathcal{A}|$ than outside, whereas for $\Ap^a$ and $\Ap^c$, comparable strengths of $|\mathcal{A}|$ are found at  locations both inside and outside the channel.

\begin{figure}
\centerline{\includegraphics[width=\textwidth]{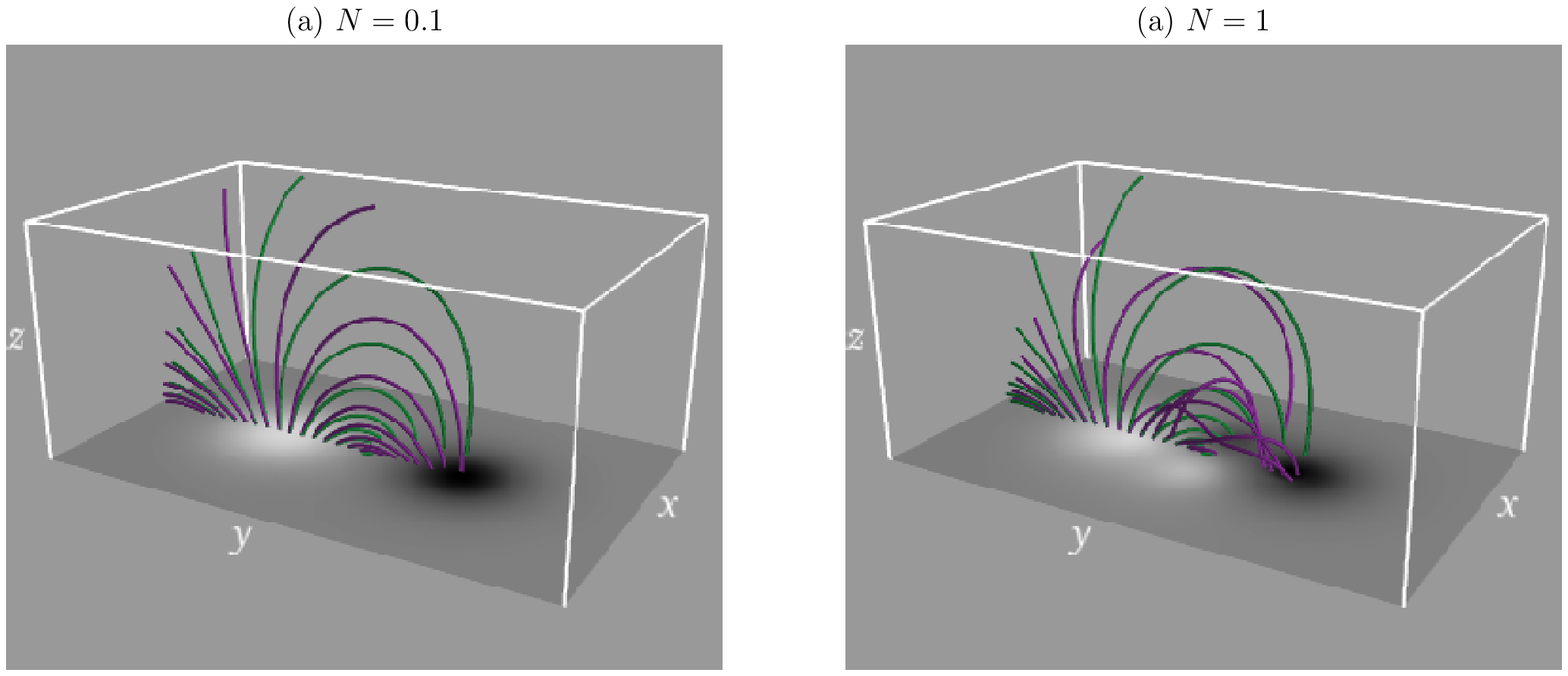}}
\caption{Field lines of $\Bb$ (purple) and $\Bp$ (green), traced from the same starting points in $y>0$ (to left in this view), for the Titov-D\'emoulin examples with (a) $N=0.1$ and (b) $N=1$.}
\label{fig:tdfl}
\end{figure}

In this example, the relative field-line helicity $\Ar^*$ (and also $\Ar^b$) is much more concentrated than in the Low-Lou equilibrium. This holds for all $N$, as shown in Figure \ref{fig:tdw}. It is strongest in the tear-drop shaped footpoints of the current channel, although diffuse non-zero values are found throughout the ``closed field'' region (i.e., the region between the diagonal separatrices within which field lines have both ends on $z=0$ and cross the central polarity inversion line). There are also thin ridges in $\Ar^*$ along the separatrices; these arise because the separatrices are in slightly different locations in $\Bb$ and $\Bp$. Outside of these separatrices, $\Ar^*$ is close to zero, reflecting the fact that these field lines are close to potential. Figure \ref{fig:tdfl} shows a comparison of field lines of $\Bb$ and $\Bp$ for two of the datacubes, showing this effect.

If we compare the maximum magnitude of $|\Ar^*|$ as $N$ is increased (Figure \ref{fig:tdw}), we see that the maximum $|\Ar^*|$ (relative to $\Phi_0$) increases from $N=0.1$ to $0.5$ and again from $N=0.5$ to $1$, but is comparable between $N=1$ and $3$, despite the fact that the maximum $J_\parallel/B_0$ increases by a factor 4. This reflects the fact that $\Ar^*$ is a global topological measure that does not solely measure the local twist (or ``self helicity'') of magnetic field lines. However, it does highlight that $|\Ar^*|$ will likely not relate directly to the stability of a given configuration, since the $N=3$ case is expected to be kink-unstable while the $N=1$ (and $N=0.1$, $0.5$) is expected to be stable \citep{2016SSRv..201..147V}. The similar peak values of $|\Ar^*|$ located in a smaller footprint for $N=3$ (as seen in Figure \ref{fig:tdw}k) is consistent with this case having a lower $|\Hr|$ than $N=1$, as found by \citet{2016SSRv..201..147V}.

\begin{figure}
\centerline{\includegraphics[width=\textwidth]{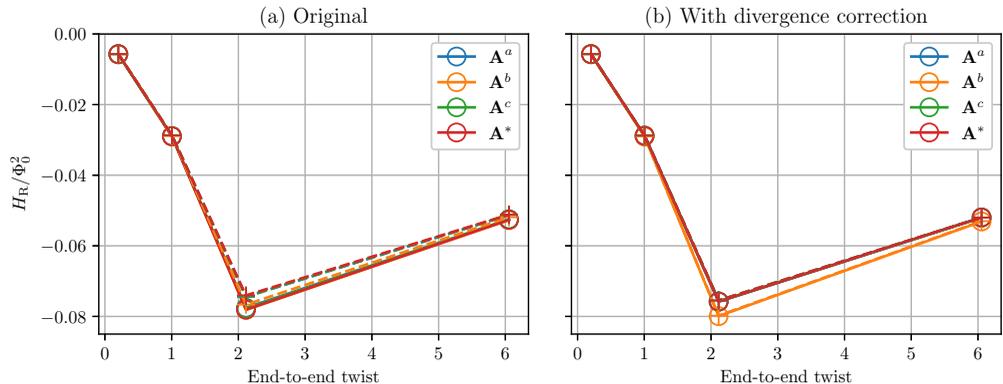}}
\caption{Numerically estimated relative helicity $\Hr$ for the Titov-D\'emoulin example, as a function of the end-to-end twist (in radians). As in Figure \ref{fig:llh}, colours denote different gauges, and solid lines/circles show direct volume-integrated $\Hr$ using the composite trapezium rule, while dashed lines/pluses are computed by integrating $\Ar|B_n|/2$ $\Ar$ over all six boundary faces. In panel (b), an additional divergence-correction step is applied, as described in the text.}
\label{fig:tdh}
\end{figure}

Finally, we repeat the validation exercise of computing $\Hr$, as for the Low-Lou model. Figure \ref{fig:tdh}(a) shows the results in the same format as Figure \ref{fig:llh}, except that the horizontal axis now labels the cases with different $N$ \citep[measured by the end-to-end twist, to facilitate comparison with Figure 3a of][]{2016SSRv..201..147V}. As for the Low-Lou example, we find consistent results when $\Hr$ is computed by volume integration. When computed from $\Ar$ via (\ref{eqn:hrfromflh}), we obtain consistent values for $\Hr$ for $N=0.1$, $0.5$ and $3$, but $|\Hr|$ is underestimated by up to $5\%$ for $N=1$, depending on the gauge. Further investigation shows that this is not due to the mesh resolution (of either the datacube or the mesh for computing $\Ar$ on the boundary), but is caused by the fact that the initial data do not perfectly satisfy the solenoidal condition on our staggered mesh. To show this, Figure \ref{fig:tdh}(b) shows the same computations of $\Hr$ when we include an additional correction to $\Bb$ after computing the (initial) $\Ab$ on  the staggered mesh. This correction recomputes $\Bb$ from the numerical curl of $\Ab$, so that the solenoidal condition is satisfied to machine precision, but at the expense of changing $\Bb$. With this correction, Figure \ref{fig:tdh}(b) shows that the estimate of $\Hr$ from $\Ar$ then agrees with the volume integral, in all four gauges. Notice however that $|\Hr|$ is then overestimated in $\Ab^b$, compared with the other gauges. This arises because the solenoidal correction has a greater effect on the current-carrying region in the downward DeVore gauge, since this region is located at lower $z$. Thus this affects the estimated helicity more. This illustrates the importance of the solenoidal condition in estimating helicity \citep{2016SSRv..201..147V}.

\section{Conclusion}

We have shown that it is possible to decompose the relative magnetic helicity $\Hr$ into a (flux-weighted) integral (\ref{eqn:hrfromflh}) over a quantity $\Ar$ called the relative field-line helicity. This quantity $\Ar(\xb)$ is itself an ideal-MHD invariant for each magnetic field line $L(\xb)$. Moreover, $\Ar(\xb)=0$ whenever $\Bb=\Bp$ all along the field line $L(\xb)$. We propose that $\Ar$ is a useful measure to identify locations of significant topological structure within a wider magnetic field. Though $\Hr$ is gauge independent, our suggested definition of $\Ar$ requires computation of a  vector potential satisfying certain conditions, as summarised in Section \ref{sec:sum}. Both the vector potential and $\Ar$ are straightforward to compute in a Cartesian domain, and we have provided free numerical code for doing so.

The principle difficulty faced in this endeavour has been the fact that $\Ar$ is not uniquely defined by the requirements of (i) ideal invariance for every field line and (ii) vanishing when $\Bb=\Bp$ all along a field line. Rather, $\Ar$ still depends on the gauge of the reference vector potential $\Ap$. Physically, we showed that this gauge dependence arises from a fundamental choice of how to define a corresponding surface for each field line. The value of $\Ar$ measures the ideal-invariant magnetic flux through this surface, but the definition of the surface depends on where it intersects the boundary, which in turn depends on the gauge of $\nb\times\Ap$ on the boundary. The gauge-dependence integrates out when integrating over all field lines to give $\Hr$, but appears to be unavoidable when considering field-line helicity alone. Nevertheless, this does not mean that $\Ar$ is physically irrelevant or useless. Rather, it means that one must choose the gauge so that the corresponding magnetic fluxes measured by $\Ar$ are informative \citep[cf.][]{2016A&A...594A..98Y}. 

\citet{2014ApJ...787..100P} discuss the physical meaning of the original $h(V)$ in different gauges, for the particular case of a magnetic field between two planes with $B_n=0$ on the side boundaries. They show that it always measures the average pairwise winding number between two curves, but with respect to a reference frame that varies depending on the gauge. They proposed the best choice of gauge to be the so-called winding gauge, $\Ab^{\rm W}$,  in which $h(V)$ measures the average winding number with respect to an untwisted Cartesian reference frame (corresponding to physical space). Here, we allow for more general magnetic fields which may enter or leave the side boundaries of the domain, so the winding gauge of \citet{2014ApJ...787..100P} does not apply. Instead, we have proposed to use a ``minimal gauge'' $\Ab^*$, also chosen by \citet{2006astro.ph..6694H}. As discussed in Section \ref{sec:astar}, such a gauge gives the simplest possible boundary distribution of $\Ab$, in a particular mathematical sense. It behaves sensibly in the examples in Section \ref{sec:eg} where it clearly peaks within the main current-carrying regions. When we consider the geometry of \citet{2014ApJ...787..100P}, $\Ab^*$ does not precisely reduce to the winding gauge $\Ab^{\rm W}$, since $\Ab^{\rm W}$ does not always satisfy condition (\ref{eqn:astar}). But preliminary computations suggest the two to be rather similar except on the side boundaries of the domain. This requires further investigation.

Finally, we have made some simplifying assumptions on the magnetic fields considered. Firstly, we have assumed that all magnetic field lines have finite length. The work of \citet{arnold1986} suggests that the field-line helicity could be extended to an asymptotic form for ergodic field lines, although the concept of field-line helicity would be useful only if a single ergodic field line does not fill the whole volume. Secondly, we have assumed a domain with simple topology. The theory could also be extended to toroidal or spherical shell volumes, but in these cases additional restrictions would be  needed on $\Ab$ so as to define $\mathcal{A}$ and $\mathcal{A}_{\rm p}$ uniquely. In fact, \citet{2016A&A...594A..98Y} have already considered the evolution of field-line helicity $\mathcal{A}$ in a spherical shell representing the global solar corona, using a specific DeVore-Coulomb gauge in spherical coordinates. Having established that this gauge was suited to the identification of magnetic flux ropes in the low corona, 
\citet{2017ApJ...846..106L} went on to use this as a tool to identify magnetic flux ropes in non-potential simulations of the global solar corona over a full solar cycle. However, no attempt was made to optimise the gauge choice in the manner of Section \ref{sec:astar}. This should also be addressed in future.

\vspace{1cm}

ARY was supported by STFC consortium grant ST/N000781/1 and Leverhulme Trust grant RPG-2017-169. MHP thanks BP for a summer studentship. The authors are indebted to G. Valori and E. Pariat for sharing the numerical data, made possible through the ISSI International Team on \textit{Magnetic helicity estimations in models and observations of the solar magnetic field}. We thank Chris Prior for suggesting improvements to an earlier draft, and the anonymous referees for further improving the paper.

\appendix
\section{Proof of minimal property}\label{app:var}
Here we prove that a vector potential satisfying (\ref{eqn:astar}) on a closed boundary $\partial V$ minimises $\oint_{\partial V}|\nb\times\Ab|^2\,\mathrm{d}^2x$ among all possible vector potentials.

To do this, suppose that $\Ab^*$ satisfies the required condition $\nabla_h\cdot\Ab^*=0$ on $\partial V$, and note that any other vector potential on $\partial V$ may be written $\Ab=\Ab^*+\nabla\phi$ for some gauge function $\phi$. Then
\begin{eqnarray}
\oint_{\partial V}|\nb\times\Ab|^2\,\mathrm{d}^2x &=& \oint_{\partial V}\Big[|\nb\times\Ab^*|^2 + 2(\nb\times\Ab^*)\cdot(\nb\times\nabla_h\phi)+ |\nabla_h\phi|^2\Big]\,\mathrm{d}^2x.
\end{eqnarray}
The cross term may be rewritten as
\begin{eqnarray}
2\oint_{\partial V}(\nb\times\Ab^*)\cdot(\nb\times\nabla_h\phi)\,\mathrm{d}^2x  &=& 2\oint_{\partial V}\nb\cdot\big(\nabla_h\phi\times(\nb\times\Ab^*)\big)\,\mathrm{d}^2x, \\
&=& 2 \oint_{\partial V}\big(\Ab^*\cdot\nabla_h\phi\big)\,\mathrm{d}^2x, \\
&=& 2 \oint_{\partial V}\nabla_h\cdot(\phi\Ab^*)\,\mathrm{d}^2x=0,
\end{eqnarray}
which vanishes because $\partial V$ is a closed surface. It follows that
\begin{eqnarray}
\oint_{\partial V}|\nb\times\Ab|^2\,\mathrm{d}^2x &=& \oint_{\partial V}|\nb\times\Ab^*|^2\,\mathrm{d}^2x + \oint_V|\nabla_h\phi|^2\,\mathrm{d}^2x,
\end{eqnarray}
so that the integral is minimised if $\nabla_h\phi=0$, i.e., if $\Ab\equiv\Ab^*$.

\bibliographystyle{jpp}

\end{document}